\documentclass[12pt]{report}
\usepackage[margin=1in]{geometry}
\usepackage{graphicx}
\usepackage{multirow}
\usepackage{natbib}
\usepackage{amssymb}
\begin{document}

\begin{titlepage}
  \begin{center}
    {\LARGE%
      Linking the Solar System and Extrasolar Planetary Systems with
      Radar Astronomy}\\
    {\Large%
      Infrastructure for ``Ground Truth'' Comparison}
  \end{center}
\noindent%
\textbf{Authors:}
Joseph Lazio ({\small Jet Propulsion Laboratory, California
  Institute of Technology}),
Amber Bonsall ({\small Green Bank Observatory}),
Marina Brozovic ({\small Jet Propulsion Laboratory, California
  Institute of Technology}),
Jon~D.~Giorgini ({\small Jet Propulsion Laboratory, California
  Institute of Technology}), 
Karen O'Neil ({\small Green Bank Observatory})
Edgard Rivera-Valentin ({\small Lunar \& Planetary Institute}), 
Anne K.~Virkki ({\small Arecibo Observatory})

\smallskip

\noindent%
\textbf{Endorsers:} Francisco Cordova ({\small Arecibo Observatory; Univ.\ Central Florida}),
  Michael Busch ({\small SETI Institute}),
  Bruce A.~Campbell ({\small Smithsonian Institution}), 
  P.~G.~Edwards ({\small CSIRO Astronomy \& Space Science}), 
  Yanga R.~Fernandez ({\small Univ.\ Central Florida}),
  Ed Kruzins ({\small Canberra Deep Space Communications Center}), 
  Noemi Pinilla-Alonso ({\small Florida Space Institute-Univ.\ Central Florida}),
  Martin A.~Slade ({\small Jet Propulsion Laboratory, California
  Institute of Technology}), 
  F.~C.~F.~Venditti ({\small Arecibo Observatory})

\smallskip

\noindent%
{\normalsize\noindent
\textbf{Point of Contact:} Joseph Lazio, 818-354-4198; Joseph.Lazio@jpl.caltech.edu}

\bigskip

{\centering
\includegraphics[width=0.33\textwidth]{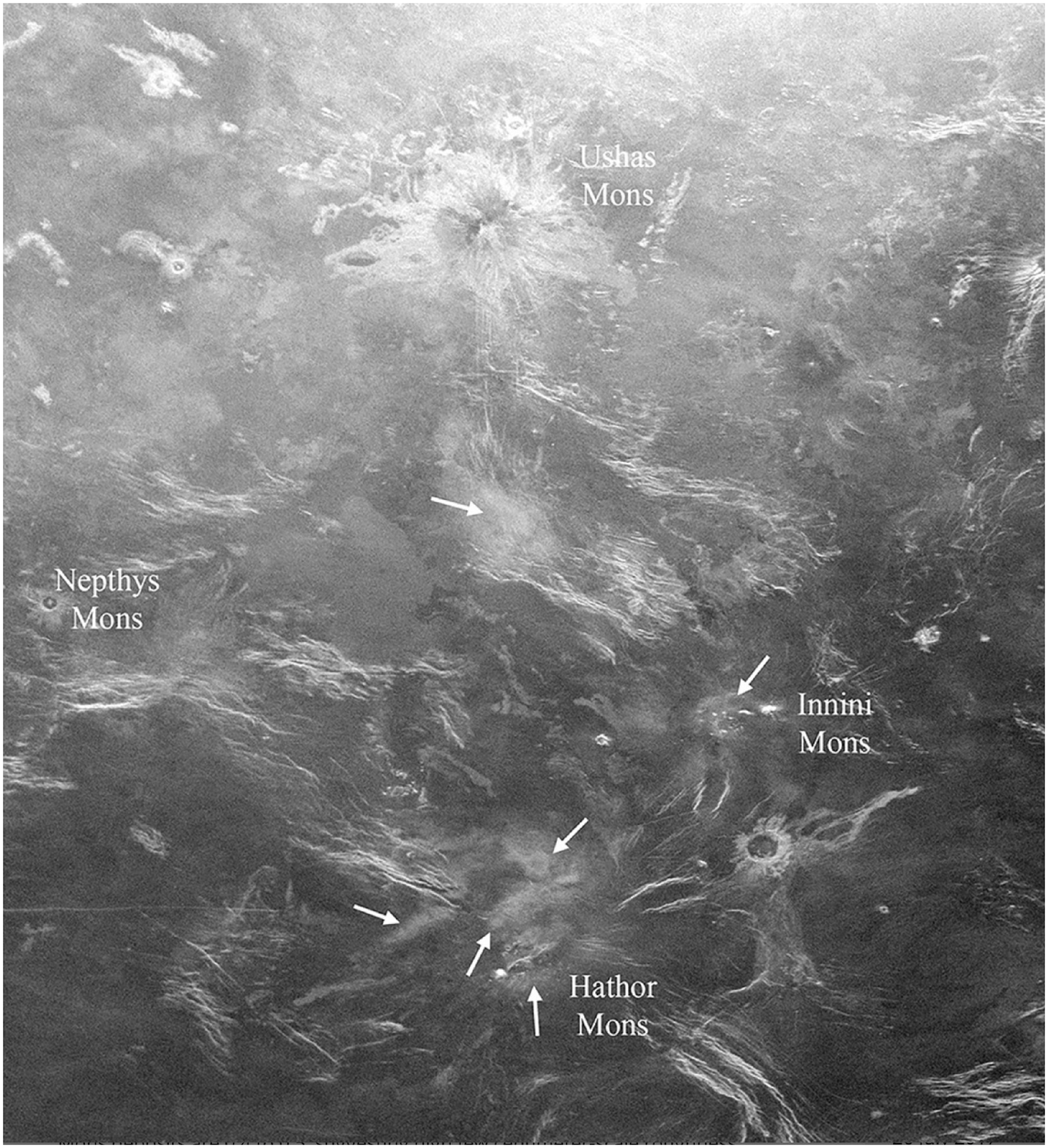}\hfil\includegraphics[width=0.57\textwidth]{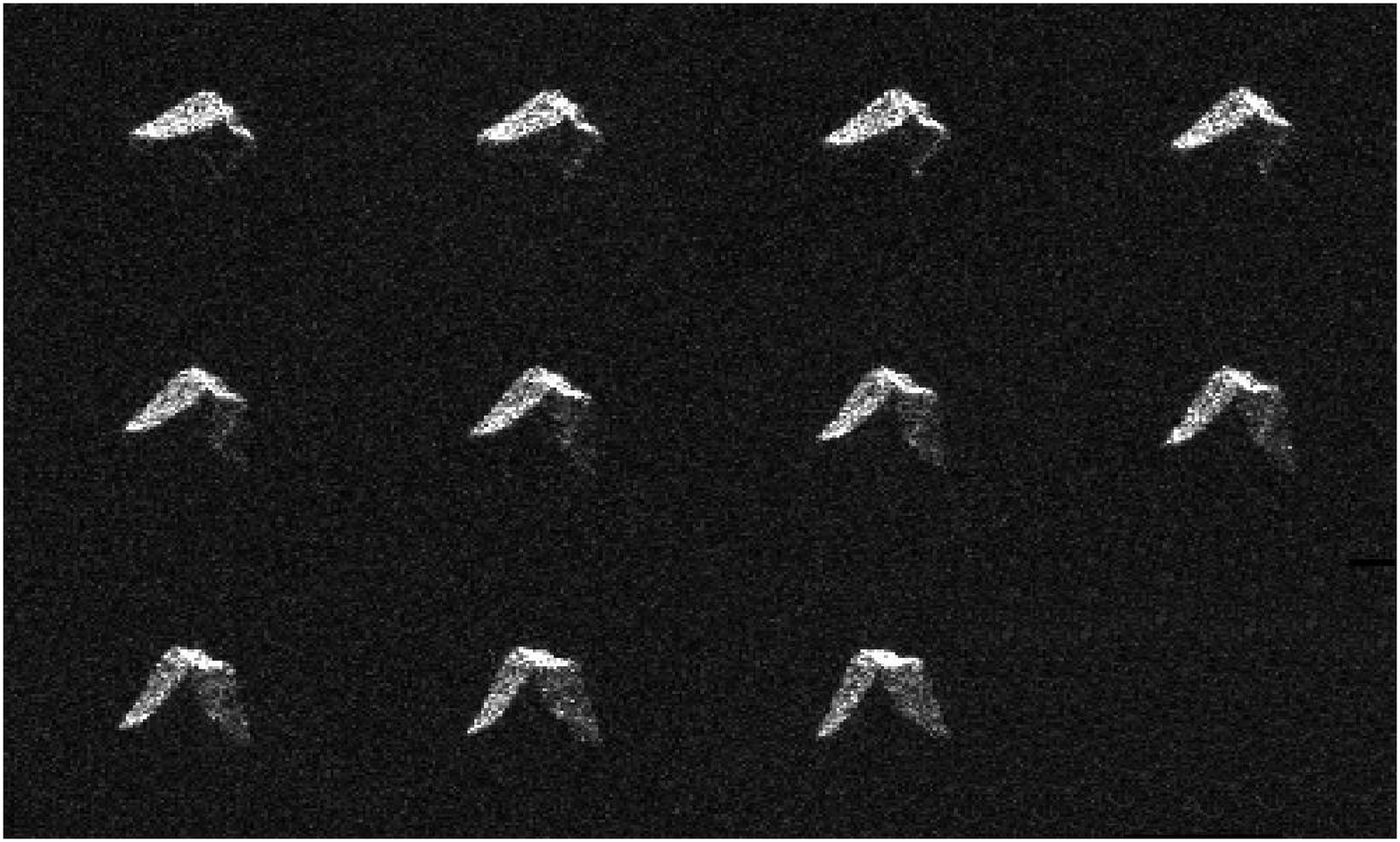}\par
\parbox{0.95\textwidth}{\textit{(Left) Planetary radar image of Dione Regio on
  Venus.  White arrows indicate radar-bright features that potentially
  represent debris from previous volcanic eruptions on Earth's twin
  planet.  \citep{cmwcgc17}
  (Right) Planetary radar image of the near-Earth asteroid
  2017~BQ${}_6$, acquired by the Goldstone Solar System Radar, showing
  sharp facets on this object.  Near-Earth asteroids
  display a range of surface features, from which evolution and
  collisional processes occurring in the Sun's debris disk can be
  constrained.}}\par
}

\bigskip

{\footnotesize\noindent%
Part of this research was carried
out at the Jet Propulsion Laboratory, California Institute of
Technology, under a contract with the National Aeronautics and Space Administration.
The Arecibo Planetary Radar Program is supported by the National
Aeronautics and Space Administration's Near-Earth Object Observations
Program through Grant No.~NNX13AQ46G awarded to Universities Space
Research Association and through grants no.~80NSSC18K1098 and
80NSSC19K0523 awarded to the University of Central Florida. The
Arecibo Observatory is a facility of the National Science Foundation
operated under cooperative agreement by University of Central Florida,
Yang Enterprises, Inc., and Universidad Ana G.~M{\'e}ndez.
The Green Bank Observatory is a facility of the National Science
Foundation and is operated by Associated Universities, Inc.  
Some of the information presented in this white paper is
pre-decisional and is for discussion and planning purposes only.}
  
\end{titlepage}

Planetary radars have obtained unique science measurements about solar
system bodies and they have provided orbit determinations allowing
spacecraft to be navigated throughout the solar system.  Notable
results have been on Venus, Earth's ``twin,'' and small bodies, which
are the constituents of the Sun's debris disk.  Together, these
results have served as ``ground truth'' from the solar system for
studies of extrasolar planets.  The Nation's planetary radar
infrastructure, indeed the world's planetary radar infrastructure, is
based on astronomical and deep space telecommunications
infrastructure, namely the radar transmitters at the Arecibo
Observatory and the Goldstone Solar System Radar, part of NASA's Deep
Space Network, along with the Green Bank Telescope as a receiving
element.  This white paper summarizes the state of this
infrastructure and potential technical developments that should be
sustained in order to enable continued studies of solar system bodies
for comparison and contrast with extrasolar planetary systems.
Because the planetary radar observations leverage existing
infrastructure largely developed for other purposes, only operations
and maintenance funding is required, though modest investments could
yield more reliable systems; in the case of the Green Bank Telescope,
additional funding for operations is required.

This white paper complements the ``GBT Planetary Radar System'' white
paper by A.~Bonsall et al.

\begin{figure}[bh]
  \centering
  \includegraphics[width=0.93\textwidth]{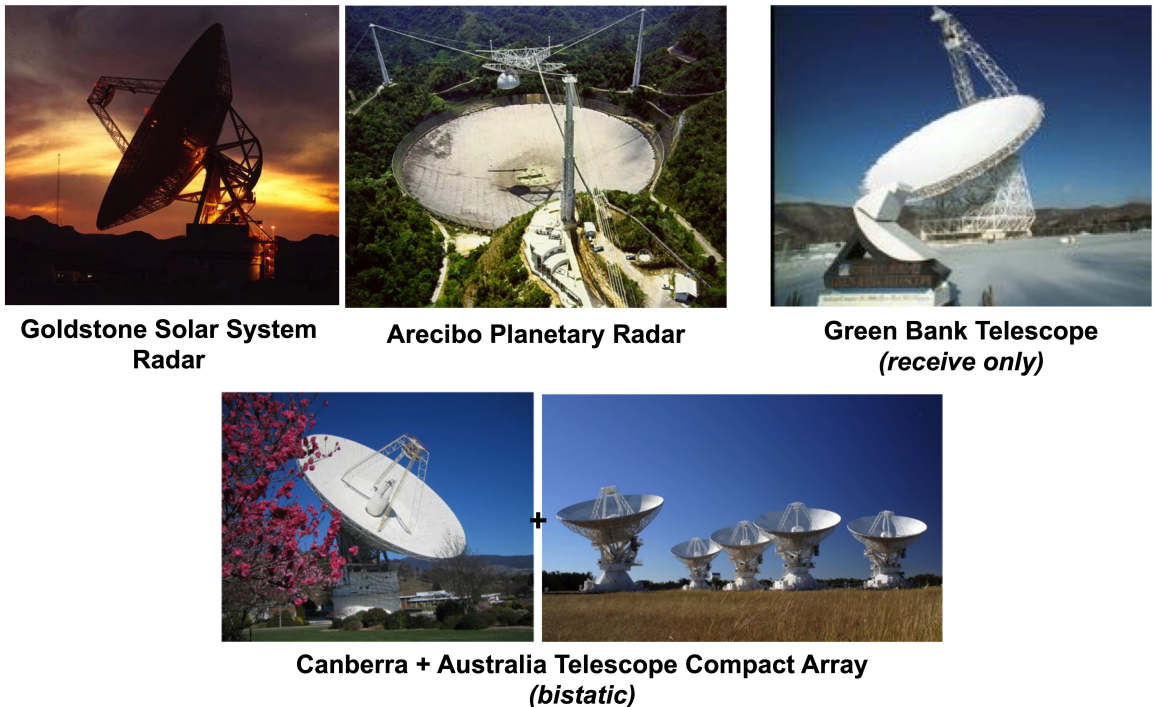}
  \vspace*{-3ex}
    \caption{International planetary radar infrastructure.
      (\textit{Clockwise from top left}) Goldstone Solar System Radar (GSSR),
    Arecibo Planetary Radar, Green Bank Telescope (\hbox{GBT}, \textit{receive
      only}), and the Southern Hemisphere system (which requires
  bistatic operations).}
  \label{fig:radar}
\end{figure}

\section*{Key Science Goals and Objectives}\label{sec:science}

Observations of solar system bodies define ``ground truth''
for interpreting observations and developing
models of extrasolar planets.  We summarize briefly some of the
linkages between radar observations of solar system bodies and
extrasolar planetary systems.  Further details are in the science
white papers by \textbf{\cite{cnc+19}}, \textbf{\cite{m+19}},
\textbf{\cite{mh+19}}, and \textbf{\cite{t+19}}.

An essential distinction for the future studies of terrestrial-mass
extrasolar planets is whether they are ``Earth-like'' or
``Venus-like'' (or ``other''!).  That these otherwise ``twin'' planets have such
different properties (atmospheric composition, surface conditions,
magnetic field strengths, \ldots) is a vivid illustration that
measurements of a planet's mass, radius, and bulk density are far from
sufficient to describe terrestrial-mass planets.  Radar observations
at centimeter wavelengths are one of the few means to probe to the
surface of Venus, due to the high atmospheric optical depth.
Moreover, planetary radar observations are vital to characterizing the
surface of Venus in a cost-effective manner; data quality is surpassed
only by dedicated missions costing in the hundreds of millions of dollars.

Planetary radar measurements provide crucial information about the
current interior properties of Venus, with implications both for
models of its formation and for the properties of extrasolar planets.
Key measurements include determining rotational variations and the
extent of vulcanism.  From rotational variations, one can constrain
the planet's moment of inertia, from which the interior mass
distribution can be inferred \citep{m+19,ccc+19}.  Standard approaches to
determining the interior mass distribution, such as measurements of
seismic activity from surface seismometers, are not feasible, and 
radar measurements are one of the few means of obtaining this
information.  The extent of vulcanism, which may be linked with
seismic activity and tectonics, is an open question.  By obtaining
global maps, and determining the extent of new surface features,
planetary radar is one of the few approaches to stringent constraints
on the extent of current volcanoes on Venus \citep{cnc+19}.

Planetary radar measurements provide precise orbit determinations and
shape and rotational state information about small bodies, most
notably asteroids \citep{mh+19,t+19}.  In the case of the solar system, the
individual constituents of the Sun's debris disk can be studied, in
contrast to the case for extrasolar planetary systems for which only
integrated properties can be determined.  The orbits and
characteristics of small bodies constrain the processes that have
shaped the Sun's debris disk, including the extent of collisions and
gravitational interactions (resonances) with the planets.

\section*{Technical Overview}\label{sec:tech}

During a planetary radar experiment, a signal with predefined 
characteristics is created and transmitted.  The signal interacts 
with and is modified by the object. The changes observed when the 
reflected signal is received on the ground are then used to make 
inferences about the object that modified the signal: its size, 
shape, spin, and reflectivity, which tells us about surface 
properties, constrains gravity, and sometimes composition.

Figure~\ref{fig:radar} and Table~\ref{tab:radar} illustrate the
international capability for conducting the desired scientific
observations; \cite{nbmbt16} provide further technical details and
comparisons between the various facilities.  Notably, most of this
observational capability is based in the United States.  The
\textbf{Goldstone Solar System Radar (GSSR)}, installed on the 70~m
Deep Space Station-14 (DSS-14) antenna at the DSN's Goldstone site,
part of NASA's Deep Space Network (DSN), and the \textbf{Arecibo
  Planetary Radar} at the 305~m-diameter Arecibo Observatory consist
of antennas equipped with both powerful transmitters and receivers.
The Robert C.~Byrd \textbf{Green Bank Telescope (GBT)} is a sensitive
receiving antenna, though it does not have a transmitter (currently).
In the \textbf{Southern Hemisphere}, there is a nascent capability,
illustrated by a combination of the \emph{bistatic} combination of the
70~m Deep Space Station-43 (DSS-43) antenna at the DSN's Canberra
Complex equipped with a 20~kW transmitter and the Australia Telescope
Compact Array (ATCA).  Not shown is the 34~m DSS-13 antenna at the
DSN's Goldstone Complex equipped with an 80~kW transmitter, which can
be used in a bistatic manner with the \hbox{GBT} and Arecibo.  For the
Southern Hemisphere, in some cases, a 34~m DSN antenna at the Canberra
Complex has been used in place of DSS-43, and early tests used the
Parkes Radio Telescope rather than the \hbox{ATCA}; however, Parkes
does not normally have a receiver mounted that can receive at the
frequencies at which DSS-43 (or other DSN Canberra Complex antennas)
can transmit.

\begin{table}[tbh]
  \centering
  \caption{Planetary Radar Infrastructure}\label{tab:radar}
{\footnotesize
  \begin{tabular}{lp{12em}cc}
    \noalign{\hrule\hrule}
\textbf{System} & \textbf{Antenna} & \textbf{Transmitter}
  & \textbf{Transmit Frequency}\\

                &                  & \textbf{Power}
  & \textbf{Wavelength}\\
    \noalign{\hrule}
\multirow{2}{*}{Arecibo Planetary Radar}      
 & \multirow{2}{10em}{305~m diameter} 
  & 900~kW 
   & 2.38~GHz \\
 &&& 12.3~cm \\
\multirow{2}{*}{Goldstone Solar System Radar}
 & \multirow{2}{10em}{70~m diameter}  
  & 450~kW 
   & 8.56~GHz \\
 &&& 3.5~cm \\
\multirow{2}{*}{Green Bank Telescope}
 & \multirow{2}{10em}{100~m diameter} 
  & \textit{500~kW}
   & \textit{32~GHz} \\
 && \textit{proposed} & \textit{1~cm} \\
\multirow{2}{*}{Southern Hemisphere}
 & \multirow{2}{11em}{70~m diameter transmit $+$50~m equivalent receive}
  & {20~kW current}
   & 7.19~GHz \\
 && \textit{80~kW (2021)} & 4.2~cm \\
    \noalign{\hrule\hrule}
  \end{tabular}
}
\end{table}

The need for high sensitivity antennas and high transmitter powers is
driven by the radar equation
\begin{equation}
\mathrm{S}/\mathrm{N}
 \propto \frac{G_{\mathrm{RX}}P_{\mathrm{TX}}G_{\mathrm{TX}}}{R^4},
  \label{eqn:radar}
\end{equation}
where S/N is the signal-to-noise ratio; $G_{\mathrm{RX}}$ and
$G_{\mathrm{TX}}$ are the gains of the receiving and transmitting
antennas, respectively; $P_{\mathrm{TX}}$ is the power of the
transmitter, and~$R$ is the range (distance) to the target.  For an
antenna of diameter~$D$, the gain is $G \propto D^2$.

This planetary radar infrastructure can observe all terrestrial
planets, and the largest moons \citep[e.g.,][]{mgbs90}, by virtue of
their large cross sections and the hundreds of kilowatts of
transmitter power coupled with antennas with the large gains (Arecibo,
\hbox{GSSR}, and the GBT).  This same infrastructure can observe
near-Earth asteroids (NEAs), with typical NEAs being several hundred
meters in diameter at ranges of order 0.1~au.  Smaller NEA targets can
be observed if their range is also smaller, with objects as small as
of order 10~m having been observed at ranges less than 1~lunar
distance.

Further, for NEA radar observations at ranges $R \lesssim 1$~lunar
distance, \textbf{bistatic} observations, even with smaller gain
antennas, can be useful to reduce stress on the transmitters.  For
nearby objects, the round-trip light travel time can be less than 1~s.
In \emph{monostatic} observations, in which the same antenna transmits
and receives, these observations would require cycling the transmitter
power (``on'' and ``off'') on sub-second time scales.  Stresses on
transmitters with powers approaching 1~MW can be significant with such
rapid power cycling, leading to premature failure or other failures.
Indeed, in the case of Arecibo, it takes 4~s (about 3 lunar distances)
to switch from the transmitter to the receiver. 

In the case of the Southern Hemisphere system, bistatic operations are
required.  The antennas at the DSN's Canberra Complex are not equipped
currently with receivers that operate at their transmitters'
frequency (approximately 7~GHz).  However, the Australia Telescope
Compact Array (ATCA) does have receivers that operate at this
frequency.

This infrastructure is complementary for its ability to observe a wide
range of targets.  With its high gain and large transmit power,
Arecibo can detect targets at the largest ranges, for a given S/N
ratio.  However, Arecibo has a limited range of declinations and can
observe only approximately 30\% of the sky (in a window centered
approximately on a declination of~$20^\circ$).  By contrast, the
\hbox{GSSR}, either monostatically or bistatically with the
\hbox{GBT}, can observe approximately 75\% of the sky.  

\section*{Technology Drivers}\label{sec:drivers}

As summarized in equation~(\ref{eqn:radar}), there are two key aspects
to a planetary radar system---transmitter power~$P_{\mathrm{TX}}$ and antenna gain~$G$.

\begin{description}
\item[Transmitter Power]%
  Fundamental to both the GSSR and Arecibo Planetary Radar are
  \emph{klystrons}---high-power, vacuum tube-based microwave
  amplifiers.  Klystrons operate by producing an electron beam that is
  then modulated to produce a radio frequency signal, often with a
  desired waveform.  While klystrons are a standard component of
  radars, some medical devices, and particle accelerators, the
  klystrons used by planetary radars are distinguished by their power
  levels, which can be orders of magnitude higher than all other
  applications.  As a consequence, they often have to be operated near
  the edge of instability, which can lead to relatively short
  lifetimes (of order a year).

  A potential technology development activity would be the development
  of even higher power but more reliable microwave amplifiers.  Both
  the GSSR and the Arecibo Planetary Radar obtain their full powers by
  the phased combinations of the outputs of two klystrons.  One likely
  approach for higher power, more reliable microwave amplifiers would
  be more modular systems, consisting of many, lower power amplifiers
  the outputs of which are combined coherently.  Lower power
  amplifiers would be presumably more reliable, and a modular approach
  would offer the possibility of a graceful degradation.  A challenge
  with such an approach would be to maintain a high efficiency of
  coherent combination of the signals, as reflected power could damage
  the individual components.  This challenge is particularly acute to
  obtain graceful degradation, as the loss of an individual amplifier
  would require a rapid response so that additional individual
  amplifiers are not damaged.

\item[Antenna Gain and Antenna Arrays]%
  Current single dish antennas are at the scale that larger antennas
  are not a viable approach to obtain significant improvements in
  antenna gain.  For instance, no fully-steerable single dish radio
  antenna significantly larger than the GBT has since been constructed
  since the completion of the GBT in the 1990s.  The continued
  operation of these large single dish antennas is therefore critical
  to a viable planetary radar capability.

  One potential approach for obtaining larger receiving
  gains~$G_{\mathrm{RX}}$ would be bistatic operations with future
  radio astronomical arrays.  For instance, in the southern
  hemisphere, the intermediate frequency
  component of the \textbf{Square Kilometre Array Phase~1 (SKA1-Mid)}
  is planned to be capable of operating at the transmitter
  frequencies of the DSN's Canberra Complex \citep{jl15}.  However, SKA1-Mid is
  planned to be sited in South Africa, while the DSN's Canberra
  Complex is in eastern Australia, limiting the amount of common sky
  visibility.  In the northern hemisphere, the \textbf{next-generation
    Very Large Array (ngVLA)} would offer both frequency coverage that
  overlaps with both the Arecibo Planetary Radar and the GSSR
  transmitters and considerable mutual visibility with both \citep{bbmnl18}.

  Finally, while interferometric (``receive-only'') arrays are
  standard for radio astronomical telescopes and phased arrays are
  standard for terrestrial radar applications, phased interferometric
  arrays are not yet used
  for the transmit portion of planetary radar.  There have been
  limited experiments within the DSN of phasing transmit antennas to
  obtain higher effective transmit gains~$G_{\mathrm{TX}}$ (the DSN
  ``Uplink Array'').  These experiments have demonstrated the expected
  improvements in $G_{\mathrm{TX}}$ for coherent addition of antennas,
  but they have not been required to obtain the range and range-rate
  (Doppler) precision required for planetary radar.  Additional work
  would be necessary to obtain the required precisions, and additional
  antennas would be necessary to obtain comparable or larger transmit
  powers.  (For example, $5 \times 34$~m antennas each equipped with
  an 80~kW transmitter $=$ 70~m antenna equipped with a 500~kW
  transmitter.)

\end{description}

\section*{Organization, Partnerships, and Current
  Status}\label{sec:org}

The planetary radar infrastructure is an effectively tripartite
organization.
\begin{description}
\item[Arecibo Observatory]%
Operates and maintains the Arecibo Planetary Radar, with funding from
\hbox{NASA}.  Proposals for radar
observations are evaluated as part of Arecibo Observatory's standard
scientific review process.  The
Arecibo Observatory itself is a facility of the \hbox{NSF}, and the
current cooperative agreement provides for a decreasing amount of
funding support from the \hbox{NSF}.  

\item[Deep Space Network]%
Operates and maintains both the GSSR and the Canberra Complex, with
funding from \hbox{NASA}.  Proposals for radar observations can be
submitted directly to the DSN or through standing elements in NASA's
Research Opportunities in Space and Earth Sciences (ROSES) omnibus
solicitation.

\item[Green Bank Observatory]%
  Operates and maintains the \hbox{GBT}, with funding from the
  \hbox{NSF}.  Proposals for radar observations are evaluated as part
  of the Green Bank Observatory's standard scientific review process.
  However, the current cooperative agreement provides for a decreasing
  amount of funding support from the \hbox{NSF}.  In order to ensure
  continued availability of the GBT for radar studies, funds should be
  allocated to dedicate GBT time toward solar system studies.
\end{description}
Within the U.{}S., there  also have been a small number of  ``radar speckle''
observations conducted with the \textbf{Very Large Array (VLA)} and
the \textbf{Very Long Baseline Array (VLBA)} used as the receiving
elements \citep{bkb+10}, both operated by the
National Radio Astronomy Observatory (NRAO).

In addition to these three organizations in the northern hemisphere,
for southern hemisphere observations, the \textbf{Australia Telescope
  National Facility (ATNF)} operates and maintains the \hbox{ATCA},
with funding support from the Australian Commonwealth Scientific and
Industrial Research Organisation (CSIRO).  The recent establishment of
an Australian Space Agency may represent additional opportunities for
support of a southern hemisphere radar system.

There is a long history of collaboration between these various
institutions.  Numerous bistatic radar observations have been
conducted, requiring coordinated scheduling of antennas. 
Further, 
the Advanced Exploration Systems (AES) Program, a joint program
between NASA's Science Mission and Human Exploration \& Operations
Mission Directorates, provided 8.56~GHz receivers for both Arecibo and
the GBT in order to enable bistatic radar observations with the GSSR
in order to enable signal-to-noise ratio improvements due to the
larger apertures of both of these antennas.  The receiver provided for
Arecibo was a copy of the GSSR digital receiver, and JPL provided the
GBT with a dual-channel analog Pinnacle Technologies ``Agile Receiver.''

Finally, as noted above, both the ngVLA and the SKA1-Mid represent
potential future opportunities for receiving elements in bistatic
radar observations.  The SKA1-Mid will be operated by the SKA Office,
and the ngVLA would be operated by the \hbox{NRAO}.  Planetary radar
observations are mentioned in the science case for both telescopes \citep{jl15,bbmnl18},
but the specific details of how bistatic observations might be
conducted with either facility have not been considered.

\section*{Schedule}\label{sec:schedule}

There are two operational systems in the northern
hemisphere (the Arecibo Planetary Radar and the GSSR) and a nascent
system in the southern hemisphere.  This state represents a change
from a decade ago when there was effectively no southern hemisphere
capability.  We focus here on potential technical changes to these
systems, leaving aside the programmatic issues noted above.

\begin{description}
\item[Early 2020s]%
  Both the Arecibo Planetary Radar and the GSSR should remain at
  approximately historical capability.  The Arecibo telescope itself
  has suffered damage due to recent hurricanes crossing the island of
  Puerto Rico, but funding is available to repair the antenna and
  restore functionality.  The GSSR is currently offline due to repair
  and upgrade of its klystrons, but the current schedule has new
  klystrons arriving and the GSSR returning to full service in~2020 to
  early 2021.

  In~2020, the DSN plans to upgrade the transmitter at DSS-43 at the
  Canberra Complex from its currently 20~kW capability to~80~kW.  This
  upgrade will take most of the year, with DSS-43 returning to service
  in~2021.  All other considerations being equal, this $4\times$
  increase in the transmitting power will increase the range to which
  the Southern Hemisphere system could detect objects by approximately
  40\% or allow it to detect objects approximately half the diameter
  it can currently.

  Also occurring in the first half of the decade are the
  investigations into alternate technologies for planetary radar,
  described above (``Technical Drivers'').

\item[Late 2020s]%
  If any of the on-going technology development proves feasible, the
  relevant technology could be implemented in the latter half of the
  2020--2030 decade.  Current plans are also for the SKA1-Mid to enter
  operations and construction to begin on the ngVLA during this interval.

\end{description}

\end{document}